\documentclass{article}

\usepackage{arxiv}
\usepackage[utf8]{inputenc} 
\usepackage[T1]{fontenc}    
\usepackage{hyperref}       
\usepackage{url}            
\usepackage{booktabs}       
\usepackage{amsfonts}       
\usepackage{nicefrac}       
\usepackage{microtype}      
\usepackage{lipsum}
\usepackage{xcolor}
\usepackage{graphicx}
\graphicspath{ {./images/} }
\usepackage{bm}

\usepackage{algorithm}
\usepackage{algpseudocode}

\usepackage{amsthm} 
\usepackage{amsmath} 
\usepackage{amssymb} 

\usepackage{dsfont}

\usepackage[round]{natbib} 

\usepackage{multirow}

\usepackage{graphicx}
\usepackage{subcaption}

\title{Regression modeling for cure factors on uterine cancer data using the reparametrized defective generalized Gompertz distribution}

\author{
 Dionisio Silva Neto \\
  Institute of Mathematics and Computer Sciences\\
  University of São Paulo\\
  São Carlos, Brazil \\ 
  and \\
Department of Statistics\\
  Federal University of São Carlos\\
  São Carlos, Brazil \\ 
  \texttt{dionisioneto@usp.br} \\
   \And
 Francisco Louzada Neto \\
  Institute of Mathematics and Computer Sciences\\
  University of São Paulo\\
  São Carlos, Brazil \\ 
  \texttt{louzada@icmc.usp.br} \\
  \And
 Vera Lucia Tomazella \\
  Department of Statistics\\
  Federal University of São Carlos\\
  São Carlos, Brazil \\
  \texttt{vera@ufscar.br} \\
}

\begin{document}
\maketitle
\begin{abstract}
Recent advances in medical research have improved survival outcomes for patients with life-threatening diseases. As a result, the existence of long-term survivors from these illnesses is becoming common. However, conventional models in survival analysis assume that all individuals remain at risk of death after the follow-up, disregarding the presence of a cured subpopulation. An important methodological advancement in this context is the use of defective distributions. In the defective models, the survival function converges to a constant value $p \in (0,1)$ as a function of the parameters. Among these models, the defective generalized Gompertz distribution (DGGD) has emerged as a flexible approach. In this work, we introduce a reparametrized version of the DGGD that incorporates the cure parameter and accommodates covariate effects to assess individual-level factors associated with long-term survival. A Bayesian model is presented, with parameter estimation via the Hamiltonian Monte Carlo algorithm. A simulation study demonstrates good asymptotic results of the estimation process under vague prior information. The proposed methodology is applied to a real-world dataset of uterine cancer patients. Our results reveal statistically significant protective effects of surgical intervention, alongside elevated risk associated with age over 50, diagnosis at the metastatic stage, and treatment with chemotherapy.
\end{abstract}


\section{Introduction}

In the recent years, many medical studies have increasingly focused on improving the quality of life of patients with severe illnesses, such as cancer and other degenerative diseases. The study by \cite{campos2009quality} reports that the majority of patients living with HIV experienced an improvement in their quality of life after initiating treatment. The authors also emphasize the relevance of psychiatric care and other complementary interventions as important contributors to enhancing and prolonging the quality of life of these patients. \cite{bourke2015survivorship} present a systematic review of randomized controlled trials evaluating supportive interventions aimed at improving the quality of life of prostate cancer patients. In this context, the term \textit{cancer survivor} is defined as the broad experience across the cancer continuum, meaning \textit{living with, through,} and \textit{beyond a cancer diagnosis}. \citet{sugimura2006long} reported that in 2005, around 170,000 individuals were diagnosed with primary lung cancer in the United States, with approximately 26,000 becoming long-term survivors each year, mainly due to improvements in treatment and surgery. \cite{vasan2019view} discuss about studies that employ more potent, targeted therapies as an initial treatment strategy, particularly those capable of retaining effectiveness despite the development of resistance mutations or preventing such mutations from emerging, can extend patient survival more effectively than delaying their use until resistance arises. The authors present a  practical example of this is seen in patients with ovarian cancer, where those who achieve a partial or complete response following platinum-based chemotherapy can experience an even more profound response with two years of maintenance therapy using olaparib medicine. Clinical outcomes showed a 60\% remission rate three years after initiating their inhibitor treatment, indicating that an increase of the proportion of patients who achieve a potential cure.


In these contexts, although the traditional survival analysis approach remains essential, it often leads to several interpretative biases.This occurs because the vast majority of survival models assume that all individuals remain at risk of experiencing the event of interest., even after a long follow-up period. Therefore, it is crucial to account for the presence of a subgroup of patients who may become cured or immune to the event of interest, which could be, for example, death or disease recurrence. A conventional approach for modeling this phenomenon is the use of standard mixture cure models, originally proposed by \cite{boag1949curefraction} and further cited by \cite{berkson1952survival}. These models define the population survival function as
$$
S_{p}(t) = (1-p) + p \, S(t),
$$
where $p \in (0, 1)$ represents the proportion of individuals who are not susceptible to the event of interest (i.e., cured or immune), and $S(t)$ denotes the parametric survival function, which can be specified using distributions such as the Weibull, log-normal, log-logistic, among others. A characteristic feature of $S_p(t)$ is its improper behavior, as it converges to $(1 - p)$ rather than zero as $t \rightarrow \infty$. The extensions of this framework have introduced regression structures to model how covariates influence the probability of cure \citep{lambert2007modeling, mazucheli2013exponentiated}. Despite its wide application, the standard mixture cure model requires the inclusion of an additional parameter ($p$) solely to capture the cure fraction, which may increase the complexity and burden of the inferential estimation process.

\cite{balka2009review} introduced the concept of defective distributions as an alternative approach to model cure fractions in mixed populations. This class of distributions naturally exhibits an improper cumulative distribution function (CDF), meaning that $F(t)$ does not integrate to one. This behavior arises by expanding the parameter space of one of the model's parameters. Consequently, due to the intrinsic relationship between the survival function and the CDF, where $S(t) = 1 - F(t)$, the survival function becomes defective naturally. This methodology offers advantages over standard mixture models by capturing the cure fraction as a function of the estimated parameters, eliminating the need to introduce an explicit additional parameter. Thus, the survival function converges to a value $p \in (0, 1)$ as $t \rightarrow \infty$, which reflects the proportion of cured individuals. This class also offers an integrated approach, as the statistical estimation proceeds in the conventional manner when there is no evidence of cure in the population.

Among defective distributions, the Gompertz distribution stands out in the literature for its natural defective behavior in scenarios involving cure fractions. Several studies have leveraged this property to propose parsimonious modeling approaches \citep{darocha2014inferencia, calsavara2019defective, toledo2023gompertz, rodrigues2024defective}. Extensions of the Gompertz distribution within the Marshall-Olkin, Kumaraswamy and Beta-G families have been explored by \cite{rocha2016twomarshall}, \cite{rocha2017new} and \cite{vieira2025defective}, respectively. Moreover, adaptations of the defective Gompertz distribution induced by a frailty term was proposed by \cite{scudilio2019defective} to take into account the latent information of non-observable heterogeneity.

In this project, we adopt the generalized Gompertz distribution, proposed by \cite{el2013generalized}, as a flexible modeling tool for continuous, non-negative random variables. This distribution exhibits defective behavior when its shape parameter assumes real values \citep{rodrigues2024defective}. Subsequently, we propose a reparameterization in terms of the cure as an estimand (i.e., a function of the estimated parameters) to interpret the defective distribution in relation to the cure fraction within the population. Additionally, we incorporate covariates into the cure parameter to explore the interpretation of risk factors and the impact of interventions on the individual probability of cure concerning the event of interest.

The rest of this paper is organized as follows: in Section \ref{sec:methodology}, we present the proposed methodology, discussing the construction of the defective regression model and the inferential process; in Section \ref{sec:Simulation Studies}, we present a simulation studies addressing different scenarios; in Section \ref{sec:applciation}, we describe the motivating dataset and apply our proposed model, followed by a diagnostic evaluation; finally, in Section \ref{sec:conclusion}, we present the conclusions and discuss directions for future research.

\section{Methodology}
\label{sec:methodology}

\subsection{The defective Gompertz distribution}

In the conventional formulation, the Gompertz distribution is defined by two strictly positive parameters: $\alpha > 0$ (the shape parameter) and $\mu > 0$ (the scale parameter). The probability density function of the Gompertz distribution for a non-negative random variable and its corresponding survival function are, respectively,

\begin{equation}
    f(t ; \alpha, \mu) = \mu e^{\alpha t} e^{-\frac{\mu}{\alpha} \left(e^{\alpha t} - 1\right)},
    \label{eq:den_Gompertz}
\end{equation}

\begin{equation}
    S(t ; \alpha, \mu) = 1 - \int_{0}^{\infty} \mu e^{\alpha t} e^{-\frac{\mu}{\alpha} \left(e^{\alpha t} - 1\right)} \,dt =  e^{-\frac{\mu}{\alpha} \left(e^{\alpha t} - 1\right)},
    \label{eq:surv_Gompertz}
\end{equation}

\noindent when the parametric space of $\alpha$ changes from $(0, \infty)$ to $(-\infty, 0)$, the distribution becomes improper (it does not integrate 1), which is connected to the idea of a cure fraction in the data ($p_{0}$) which can be easily computed as 

\begin{equation}
    \label{eq:basal_gompertz_cf}
    p_{0} = \lim_{t \rightarrow \infty}  S(t ; \alpha, \mu) =  \lim_{t \rightarrow \infty} e^{-\frac{\mu}{\alpha} \left(e^{\alpha t} - 1\right)} = e^{\frac{\mu}{\alpha}} \in (0,1).
\end{equation}

The above expression implies that, once the values of $\alpha$ and $\mu$ are estimated, it is possible to obtain the proportion of cure fraction in data. This is a substantial advantage of this distribution compared to the traditional mixture cure model because the proportion of cure is intrinsically obtained from the survival function. This reduction of the components in the parametric vector leads to several benefits, such as improved estimation precision and simpler computations. Additionally, the model can be specified in an integrated way, where $\alpha \in \mathbb{R}$, this means the cure fraction can be detected naturally depending on the value of the shape parameter.


\subsection{The defective generalized Gompertz distribution}

To introduce greater flexibility to the defective Gompertz distribution, we work with its inclusion within the family of flexible distributions defined by \cite{lehmann1953power}, given by the cumulative distribution function:

\begin{equation}
    G(z ; \theta, \psi) = \mathbb{P}(Z \leq z) = \left[F(z \, ; \theta)\right]^{\psi},
\end{equation}

\noindent where $F(z ; \theta)$ denotes the CDF of a random variable $Z$, indexed by the parameter vector $\theta$. The parameter $\psi > 0 $, known as the \textit{power parameter}, introduces additional flexibility to the baseline distribution by modifying its shape. When $\psi = 1$, the baseline model is recovered as a particular case.

We can easily de derive the probability density and survival functions in the Lehmann's family, respectively,

\begin{equation}
    \label{eq:den_L}
    g(z ; \theta, \psi) = \psi \, f(z  \, ; \theta) \, \left[F(z  \, ; \theta)\right]^{\psi-1},
\end{equation}

\begin{equation}
    \label{eq:surv_L}
    S_{G}(z ; \theta, \psi) = 1 - \left[F(z  \, ; \theta)\right]^{\psi}.
\end{equation}

Based on Equations \ref{eq:den_L} and \ref{eq:surv_L}, applied to the probability density and survival functions of the Gompertz distribution, given in Equations \ref{eq:den_Gompertz} and \ref{eq:surv_Gompertz}, respectively, we obtain the following expressions for the probability density function and the survival function of the generalized Gompertz distribution:

\begin{equation}
    f(t ; \alpha, \beta, \psi) = \psi \, \beta e^{\alpha t} e^{-\frac{\beta}{\alpha} \left(e^{\alpha t} - 1\right)} \, \left[1 - e^{-\frac{\beta}{\alpha} \left(e^{\alpha t} - 1\right)}\right]^{\psi - 1},
    \label{eq:density_E_Gompertz}
\end{equation}

\begin{equation}
    S(t ; \alpha, \beta, \psi) = 1 - \left\{1 -e^{-\frac{\beta}{\alpha} \left(e^{\alpha t} - 1\right)}\right\}^{\psi},
    \label{eq:survival_E_Gompertz}
\end{equation}

\noindent where $\alpha > 0$, $\beta > 0$, $\psi > 0$, and $t > 0$. This family of distributions was initially discussed by \cite{el2013generalized} and later extended into a broader class of distributions presented in \cite{rocha2017new}.

A relevant question concerns whether the generalized Gompertz distribution distribution can present a defective form and under which conditions this characteristic may occur. This distribution can be seen as a particular case of the defective distribution introduced by \cite{rocha2017new}. When one of the added parameters in the Kumaraswamy family takes the value 1, the model simplifies to the baseline structure discussed in this work. We opted not to adopt the most general version of the model, which incorporates two shape parameters, in order to achieve greater flexibility while preserving the principle of parsimony.

The generalized Gompertz distribution is defective when $\alpha < 0$ (as is the case with the defective Gompertz distribution), the resulting model corresponds to the defective generalized Gompertz distribution (DGGD) as discussed by \cite{rodrigues2024defective}. Given $\alpha < 0$, the proportion of cure fraction can be easily computed as

\begin{align}
    p &= \lim_{t \rightarrow \infty} S(t ; \alpha, \beta,  \psi) = \lim_{t \rightarrow \infty}  1 - \left\{1 -e^{-\frac{\beta}{\alpha} \left(e^{\alpha t} - 1\right)}\right\}^{\psi} \nonumber \\
    &=  1 - \left\{1 - \lim_{t \rightarrow \infty} e^{-\frac{\beta}{\alpha} \left(e^{\alpha t} - 1\right)}\right\}^{\psi} = 1 - \left\{1 - p_{0}\right\}^{\psi}.
\end{align}

\noindent where $p_0=e^{\mu/\alpha}$ is the computed cure fraction in the Gompertz defective distribution (Equation \ref{eq:basal_gompertz_cf}).

\subsection{Model with covariates}

In this section, we introduce the extension of the DGGD which incorporates covariates and reparametrization. Our formulation aim to interpret the model in terms of the cure, we propose a reparametrized version in terms of the estimand $p$,

$$
p = 1-[1-p_0]^{\psi} = 1-\left[1- e^{\frac{\mu}{\alpha}}\right]^{\psi}.
$$

We propose substituting the scale parameter to maintain the identification of cure fraction in terms of $\alpha$ and the flexibility in terms of $\psi$,

$$
p = 1 - [1- e^{\frac{\beta}{\alpha}}]^{\psi} \Longleftrightarrow \mu = \alpha \ln(1-[1-p]^{\frac{1}{\psi}}).
$$

The new versions for the probability density and survival functions of the DGGD are, respectively, 

\begin{equation}
    f(t ; \alpha, p, \psi) = \psi \,\alpha \ln(1-[1-p]^{\frac{1}{\psi}}) e^{\alpha t} e^{- \ln\left(1-[1-p]^{\frac{1}{\psi}}\right) \left(e^{\alpha t} - 1\right)} \, \left[1 - e^{-\ln\left(1-[1-p]^{\frac{1}{\psi}}\right) \left(e^{\alpha t} - 1\right)}\right]^{\psi - 1} ,
    \label{eq:density_K_Gompertz}
\end{equation}
    
\begin{equation}
    S(t ; \alpha, p, \psi) = 1 - \left\{1 - e^{-\ln\left(1-[1-p]^{\frac{1}{\psi}}\right) \left(e^{\alpha t} - 1\right)}\right\}^{\psi}.
    \label{eq:survival_K_Gompertz}
\end{equation}

We include the covariate effects ($\mathbf{x}_{i}, i = 1, 2, \dots, n$), where $\mathbf{x}_{i}$ is the covariate information for the $i$-the observation, in the analysis to the cure parameter $p$ to understand how clinical characteristics can influence the probability of being cured. Considering the parametric space of $p$, we apply the logistic link function

$$
p(\mathbf{x}_{i}) = \dfrac{e^{\mathbf{x}_{i}^{\top}\boldsymbol{\beta}}}{1+e^{\mathbf{x}_{i}^{\top}\boldsymbol{\beta}}} = \dfrac{1}{1+e^{-\mathbf{x}_{i}^{\top}\boldsymbol{\beta}}},
$$

\noindent where $\mathbf{x}_{i}^{\top} = (1, {x}_{i1}, {x}_{i2}, \dots, {x}_{iq})^{\top}$ is a vector of observations from $q$ independent variables for the $i$-th observation and $\boldsymbol{\beta} = (\beta_{0}, \beta_{1}, \dots, \beta_{q})$ is the vector of regression coefficients.

\subsection{Bayesian inference}

The Bayesian paradigm focuses on obtaining the joint posterior distribution $\pi(\boldsymbol{\vartheta} \mid D)$, which results from combining prior knowledge (or ignorance) about the parameter vector $\boldsymbol{\vartheta}$, expressed through the prior distribution $\pi(\boldsymbol{\vartheta})$, with the information provided by the data via the likelihood function $L(\boldsymbol{\vartheta} \mid D)$. The posterior distribution is obtained by applying the Bayes' theorem

$$
\pi(\boldsymbol{\vartheta}\mid D) = \dfrac{\pi(\boldsymbol{\vartheta}) \times L(\boldsymbol{\vartheta} \mid D)}{\pi(D)} \propto \pi(\boldsymbol{\vartheta}) \times L(\boldsymbol{\vartheta} \mid D),
$$

\noindent where "$\propto$" denotes the proportional information about $\boldsymbol{\vartheta}$. 

Consider an independent sample of size $n$ and the observed time is $T_i = \min(T^{*}_i, C_i)$, where $T^{*}_i$ is the true survival time, $C_i$ the censoring time, with $\delta_i = \mathbb{I}(T^{*}_i \leq C_i)$ being the failure indicator and let $\mathbf{x}_{i} = (1, {x}_{i1}, {x}_{i2}, \dots, {x}_{iq})$ be the observed variables which affect the individual cure rate for the $i$-th observation. Thus, the complete observed dataset is $D = \left\{(T_{i}, \delta_{i}, \mathbf{x}_{i}; i = 1, \dots, n)\right\}$. The vector of parameters is given by $\boldsymbol{\vartheta} = (\alpha, \psi, \boldsymbol{\beta})$. The likelihood function for right-censored  survival data can be expressed as 

\begin{align*}
    L(\boldsymbol{\vartheta} \mid D) &= \prod_{i=1}^{n}  f(t ; \alpha, p_{i}, \psi)^{\delta_{i}} \times S(t ; \alpha, p_{i}, \psi)^{(1-\delta_{i})} \\
                            &= \prod_{i=1}^{n} \left[\psi \,\alpha \ln(1-[1-p_{i}]^{\frac{1}{\psi}}) e^{\alpha t} e^{- \ln(1-[1-p_{i}]^{\frac{1}{\psi}}) \left(e^{\alpha t} - 1\right)} \, \left[1 - e^{-\ln(1-[1-p_{i}]^{\frac{1}{\psi}}) \left(e^{\alpha t} - 1\right)}\right]^{\psi - 1}\right]^{\delta_{i}} \\
                            & \quad \quad \times \left[ 1 - \left\{1 - e^{-\ln(1-[1-p_{i}]^{\frac{1}{\psi}}) \left(e^{\alpha t} - 1\right)}\right\}^{\psi}\right]^{(1-\delta_{i})}.
\end{align*}

 Assuming independence among the components of the vector $\boldsymbol{\vartheta}$, the prior distribution is

$$
\pi(\boldsymbol{\vartheta}) = \pi(\alpha) \, \pi(\psi) \, \prod_{k=0}^{q} \pi(\beta_{k}), 
$$

where each component in the Bayesian regression defective generalized Gompertz model has the following prior specifications

    $$
    \beta_{k} \sim Normal(\mu_k, \sigma^{2}_{k}), k=0, 1, 2, \dots, q.
    $$

    $$
    \alpha \sim Normal(m, s^2),
    $$

    $$
    \psi \sim gamma(\gamma, \omega),
    $$

\noindent where $(\mu_k, \sigma_{k})$, $k = 0, 1, 2, \dots, p$; $m$, $s$, $\gamma$ are $\omega$ are known hyperparameters.

However, the joint posterior density \( \pi(\boldsymbol{\vartheta} \mid D) \) usually involves necessary integrals that are not easy to calculate making it impossible to obtain a closed-from. To overcome this, the Markov Chain Monte Carlo (MCMC) methods is used to obtain samples from \( \pi(\boldsymbol{\vartheta} \mid D) \), which enables inference on the parameter vector \( \boldsymbol{\vartheta} \) trough summary statistics. Two famous algorithms to sample from the posterior distribution are Metropolis-Hastings \citep{metropolis1953equation, hastings1970monte} and Gibbs sampling \citep{geman1984stochasticgiibs}. However, a major challenge in these approaches lies in selecting an appropriate proposal density to ensure efficient sampling. One interesting alternative is the Hamiltonian Monte Carlo (HMC) \citep{neal2011mcmc,duane1987hybrid}, which is based on gradients of the objective distribution to propose samples adaptively, available in \texttt{rstan} package \citep{rstancitation}.

The idea of HMC uses concepts of gradients and differential geometry on the posterior distribution. \cite{betancourt2017geometric} provide a full description of the algorithm. We can describe briefly in the following steps:

\begin{itemize}
    \item [($i$)] The state space is augmented by momentum parameters, therefore, the parameter vector consists of the parameters of interest and the momentum parameters;
    \item [($ii$)] We define the Hamiltonian function as the negative value of logarithm of the joint distribution with all parameters;
    \item [($iii$)] The momentum of all parameters is sampled from a multivariate gaussian, typically from the current value of the parameters;
    \item [($iv$)] The proposal distribution of parameters of interest is defined conditioned of the gradients of the Hamiltonian function in the current value, then, we consider the local geometry of the distribution.
\end{itemize}

The standard version of the HMC algorithm involves a large number of hyperparameters, which complicates the automation of the sampling process. These include the number of updates performed before the acceptance or rejection step (leapfrog steps), the step size (which follows the direction of the computed gradient), and the covariance matrix of the probability distribution for the momentum parameters. In \texttt{rstan} package, an adaptive version of leapfrog step is implemented to reduce the quantity of hyperparameters during tuning. The covariance matrix of momentum parameters is estimated during the warm-up period, as is the step-size, with the objective of improving the acceptance rate. The optimal number of updates is determined dynamically. The idea is to use the enough number of updates to explore the parametric space efficiently.  This occurs because the algorithm avoids retracing previously explored paths (U-turns) or halts the trajectory once a predefined maximum number of leapfrog steps is reached. The NUTS algorithm implemented in \texttt{rstan} employs multinomial sampling across the generated trajectory to select a sample \citep{hoffman2014no, betancourt2017geometric}. If the leapfrog integration process fails — indicated by a significant deviation of the Hamiltonian value from its starting point — the trajectory is classified as divergent and subsequently discarded.

Compared to conventional MCMC approaches, HMC demands greater computational resources per iteration, primarily due to the need for gradient evaluations. Nevertheless, this characteristic allows HMC to navigate complex posterior distributions with strong parameter correlations more efficiently than traditional MCMC methods. As a result, fewer iterations are generally required to obtain reliable parameter estimates and credible intervals, often leading to reduced overall computation time. Notably, \cite{monnahan2017faster} showed that across various applications, HMC implemented in \texttt{rstan} consistently achieves a higher effective sample size per unit of computation than MCMC algorithms, such as those available in JAGS (Just Another Gibbs Sampler) software.

\subsection{Residual analysis}

In order to verify the error assumptions and the presence of outliers, we apply two types of residuals analysis for our defective model for both inferences: a deviance component residual and a martigale-type residual. The martingale residuals are skew, they have maximum value +1 and minimum
value $-\infty$. In the parametric survival models, the martingale residuals can be expressed as

$$
r_{Mi} = \delta_{i} + \log(S(t_{i}; \boldsymbol{\vartheta})), 
$$

\noindent where $S(t_{i}; \boldsymbol{\vartheta})$ is the survival function computed in sample data under the vector of parameters estimated ($\boldsymbol{\vartheta}$), and $\delta_{i}$ is the indicator function for the occurrence of event of interest.

The deviance component residual is a transformation of martingale residuals to lessen its skewness, this approach is the same for generalized linear models \citep{therneau1990martingale}. In particular, the deviance component residuals to parametric regression model with explainable covariates can be expressed
as

$$
r_{Di} = \text{signal}(r_{Mi})\{ -2[r_{Mi} + \delta_{i} \, \log(\delta_{i} - r_{Mi})] \}^{-1/2}
$$

\noindent where $r_{Mi}$ is the martigale residual.

\section{Simulation}
\label{sec:Simulation Studies}

In this section, we run a simulation study to evaluate consistency through frequentist properties of the Bayesian regression model using point and interval estimates. Let $F(t)$ represent the cumulative function for the survival time associated with a specified event of interest, which can contain information of long-term survivors. Our goal is to generate the information of the observed survival time, the censoring indicator and covariates information under a stochastic process. This procedure is clearly explained in the Algorithm \ref{algo1}.

\begin{algorithm}
\caption{Dataset generation algorithm from the DGGD with covariates.}
\begin{algorithmic}[1]
\State Define the values of $\boldsymbol{\beta} = (\beta_0, \beta_1, \beta_2) \in \mathbb{R}^3$, $\alpha<0$, and $\psi > 0$;
\For{$i = 1$ to $n$}
    \State Define $\boldsymbol{x}_{i}$ = (1, ${x}_{i1}, {x}_{i2}$), where $x_{i1} \sim \text{Bernoulli}(0.5)$, $x_{i2} \sim \text{Normal}(0, 1)$;
    \State Determine the individual cure rate $$p_i(x_{i1}, x_{i2}) = \dfrac{1}{1 + e^{-\boldsymbol{x}_{i}^{\top}  \boldsymbol{\beta}}};$$
    \State Generate $M_i \sim \text{Bernoulli}(1 - p_i(x_{1}, x_{2}))$;
    \If{$M_i = 0$}
        \State Set $t_i^* = \infty$;
    \Else
        \State Take $t_i^*$ as the root of $F(t) = u$, where $u \sim \mathcal{U}(0, 1 - p_i(x_{i1}, x_{i2}))$ and $F(t)$ is the CDF of the defective generalized Gompertz distribution;
    \EndIf
    \State Generate $u_i^* \sim \mathcal{U}(0, \max(t_i^*))$, considering only the finite values of $t_i^*$;
    \State Calculate $t_i = \min(t_i^*, u_i^*)$ and $\delta_i = 1(t_i^* \leq u_i^*)$;
\EndFor
\State \textbf{Output:} The final dataset is $D = \{(t_i, \delta_i, x_{1i}, x_{2i}) : i = 1, 2, \ldots, n\}$.
\end{algorithmic}
\label{algo1}
\end{algorithm}

We evaluated four samples sizes ($n = 100, 300, 500, 1000$) over 1000 Monte Carlo replicates. In each iteration, we computed the posterior means, posterior standard deviations, relative bias, given by $B(\hat{\boldsymbol{\vartheta}}) = {(\hat{\boldsymbol{\vartheta}} - \boldsymbol{\vartheta})/\boldsymbol{\vartheta}} \times 100$, where $\hat{\boldsymbol{\vartheta}}$ is the vector of posterior means, and the function to check if each component of $\boldsymbol{\vartheta}$ is covered by the 95\% credibility intervals.  

The results show that for a sample size of $n = 300$, the mean estimates of the parameters were already close to the true values, and low variability is observed through the average values of the standard deviations. The relative bias shows satisfactory values when $n = 500$, as does the coverage probability with respect to the 95\% credibility intervals, suggesting that the model provides well-calibrated uncertainty estimates under moderate sample sizes. For the largest sample size explored ($n = 1000$), the results remain consistent, presenting adequate values across the metrics even under low prior information.

\begin{table}[]
\centering
\caption{Averages of posterior means, posterior standard deviations, relative bias and coverage probability for simulated data from the defective generalized Gompertz model.}
\begin{tabular}{ccccccc} \\ \hline
N                     & Parameter & True & Mean    & SD     & Bias\%  & Coverage Probability \\ \hline
\multirow{5}{*}{100}  & $\beta_0$     & -1.0   & -1.6385 & 0.9162 & 63.8503 & 0.9240               \\
                      & $\beta_1$     & 0.5  & 0.6569  & 0.3541 & 31.3881 & 0.9200               \\
                      & $\beta_2$     & 0.5  & 0.6517  & 0.6158 & 30.3468 & 0.9350               \\
                      & $\alpha$  & -2.0   & -1.9623 & 0.6088 & -1.8856 & 0.9100               \\
                      & $\psi$    & 2.0    & 2.2374  & 0.6603 & 11.8716 & 0.9220               \\ \hline
\multirow{5}{*}{300}  & $\beta_0$     & -1.0   & -1.0460 & 0.2311 & 4.6012  & 0.9460               \\
                      & $\beta_1$     & 0.5  & 0.5082  & 0.1356 & 1.6346  & 0.9420               \\
                      & $\beta_2$     & 0.5  & 0.5100  & 0.2571 & 1.9989  & 0.9500               \\
                      & $\alpha$  & -2.0   & -2.0064 & 0.2815 & 0.3192  & 0.9370               \\
                      & $\psi$    & 2.0    & 2.0756  & 0.3067 & 3.7808  & 0.9320               \\ \hline
\multirow{5}{*}{500}  & $\beta_0$     & -1.0   & -1.0170 & 0.1671 & 1.7045  & 0.9490               \\
                      & $\beta_1$     & 0.5  & 0.5168  & 0.1027 & 3.3565  & 0.9480               \\
                      & $\beta_2$     & 0.5  & 0.4955  & 0.1952 & -0.8942 & 0.9470               \\
                      & $\alpha$  & -2.0   & -2.0119 & 0.2059 & 0.5941  & 0.9590               \\
                      & $\psi$    & 2.0   & 2.0464  & 0.2282 & 2.3194  & 0.9420               \\ \hline
\multirow{5}{*}{1000} & $\beta_0$     & -1.0   & -1.0109 & 0.1131 & 1.0884  & 0.9520               \\
                      & $\beta_1$     & 0.5  & 0.5046  & 0.0712 & 0.9103  & 0.9570               \\
                      & $\beta_2$     & 0.5  & 0.5054  & 0.1358 & 1.0835  & 0.9460               \\
                      & $\alpha$  & -2.0   & -2.0043 & 0.1368 & 0.2163  & 0.9390               \\
                      & $\psi$    & 2.0    & 2.0185  & 0.1553 & 0.9235  & 0.9520         \\ \hline      
\end{tabular}
\end{table}

\section{Motivating data}
\label{sec:applciation}

The dataset used in this study was provided by the Oncocenter Foundation of São Paulo (FOSP), an institution linked to the São Paulo State Department of Health. The FOSP organizes clinical data from both public and private healthcare institutions across the state of São Paulo, which is systematically compiled within the Hospital Cancer Registry (HCR). The HCR includes information on cancer cases reported since the early 2000s, encompassing patients with or without a confirmed diagnosis, including individuals already undergoing treatment. In addition, the registry contains sociodemographic data and variables related to healthcare provision, providing a comprehensive overview of the oncology care profile. Using data from the HCR, we investigated cases of uterine cancer among women residing in the state of São Paulo between 2012 and 2020. 

Uterine cancer can be broadly classified into two main subtypes: endometrial cancer and uterine sarcoma. Endometrial cancer originates in the inner lining of the uterus (the endometrium) and accounts for the vast majority of uterine cancer cases. Acoording to \cite{pather2007endometrial}, it is the fourth most frequently diagnosed cancer among women, following breast, colorectal, and lung cancer, and represents the seventh leading cause of cancer-related mortality in women.
The highest prevalence is observed among postmenopausal, middle-aged, elderly, or obese women. The incidence of endometrial cancer is particularly high in western countries and shows a strong association with rising obesity rates \citep{pather2007endometrial}. By the end of the 20th century, approximately 42,000 deaths were attributed to uterine cancer worldwide, with 27,500 occurring in developed countries and 14,400 in developing regions in 1990. For instance, in the United States, mortality rates from endometrial cancer doubled between 1988 and 1998, a trend likely driven by increased life expectancy and rising obesity rates, along with associated comorbidities \citep{jemal2004cancer, parkin1999global}. In contrast, uterine sarcomas are rare tumors developed in the myometrium (the muscular layer of the uterus) or in other supporting tissues. They represent approximately 1\% of all female genital tract malignancies and 3\% to 7\% of uterine cancers \citep{major1993prognostic}. Despite their rarity, uterine sarcomas exhibit highly aggressive behavior. Among these, carcinosarcomas are especially notable for their poor prognosis, with an estimated five-year overall survival of approximately 30\%, increasing to about 50\% for patients diagnosed at stage I, when the disease is confined to the uterus. The rarity and histopathological heterogeneity of uterine sarcomas have contributed to a lack of consensus regarding optimal treatment strategies and risk factors for poor outcomes \citep{giuntoli2003retrospective}.

\begin{figure}[h]
    \centering
    \includegraphics[width=0.45\linewidth]{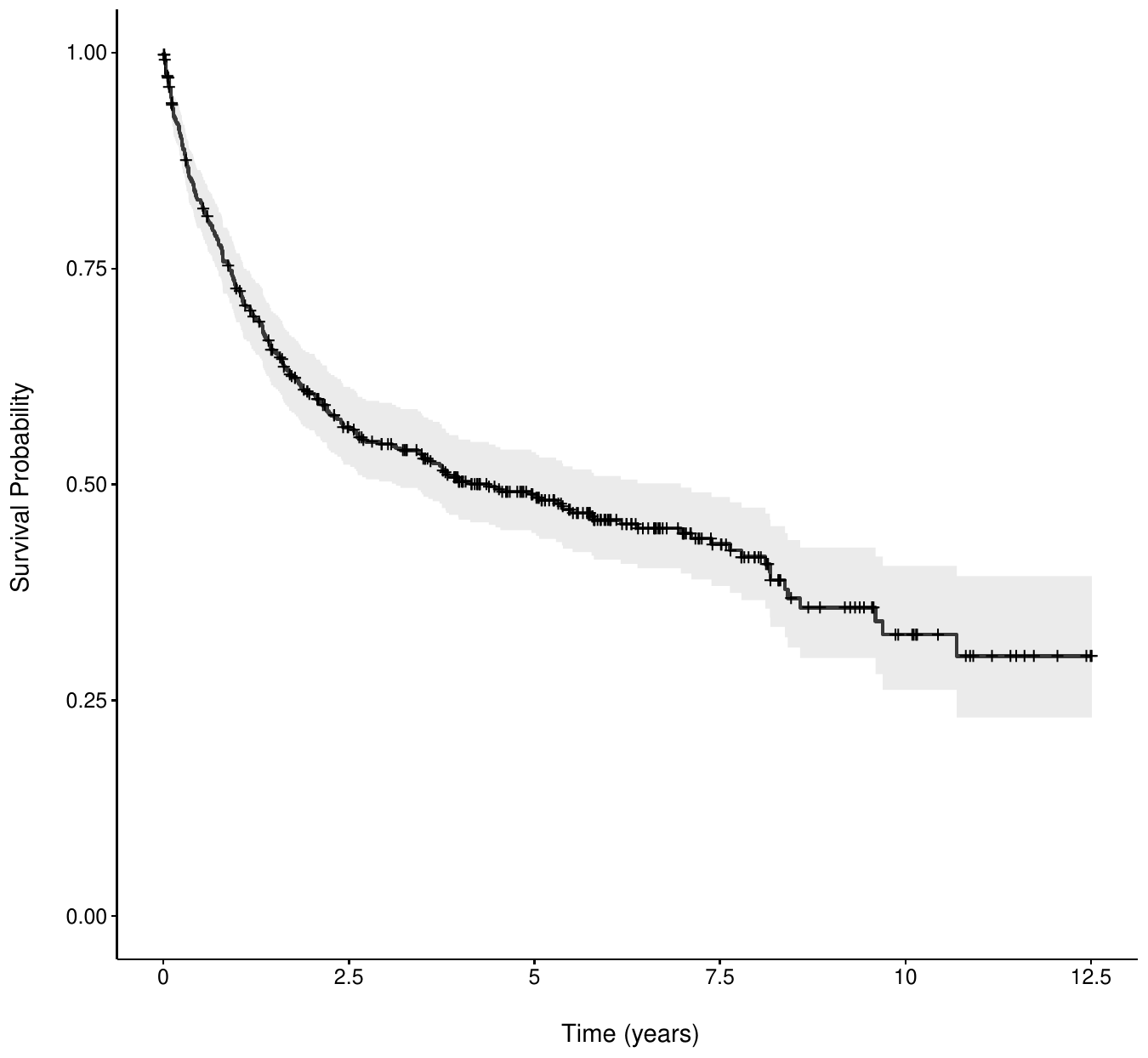}
    \caption{Kaplan-Meier curve of patients with uterine cancer.}
    \label{fig:kaplan_meier_curve_utero}
\end{figure}

In our study, we considered the failure time as the period between the date of the patient's diagnosis of uterine cancer and the date of the last available information regarding their health status (cancer-free, living with uterine cancer, or death due to uterine cancer or other causes). The event of interest was defined based on death attributed exclusively to uterine cancer. Figure \ref{fig:kaplan_meier_curve_utero} presents the nonparametric Kaplan-Meier estimator for the uterine cancer dataset, along with the 95\% confidence interval. A plateau is observed after approximately 10 years of patient follow-up, suggesting the potential presence of cured individuals within the population.

To model the probability of cure, we included the following covariates: patient age categorized into two groups, distinguishing individuals over 50 years old ($X_1$); presence of distant recurrence, representing the occurrence of metastasis ($X_2$); whether the patient underwent surgery inside or outside the hospital ($X_3$); whether the patient received chemotherapy inside or outside the hospital ($X_4$); and whether the patient received hormone therapy inside or outside the hospital ($X_5$). We adopted vague prior distributions for the regression coefficients, as well as for the shape and power parameters of the DGGD, following the same specifications described in the simulation study in Section \ref{sec:Simulation Studies}.

\begin{table}[h]
\centering
\caption{Estimates values of mean, standard deviation (SD) and 95\% credibility interval for uterine cancer data using the DGGD.}
\begin{tabular}{lccc} \hline
Parameter   & Mean    & SD     & C.I (95\%)            \\ \hline
$\beta_{0}$ & -1.9647 & 1.7328 & {[}-6.7359 ; -0.3247{]}  \\
$\beta_{1}$ & -1.6007 & 0.7747 & {[}-3.7989 ; -0.7160{]} \\
$\beta_{2}$ & -1.8502 & 1.4523 & {[}-5.2649 ; -0.0252{]} \\
$\beta_{3}$ & 2.3608  & 1.3532 & {[}1.0734 ; 6.2822{]}   \\
$\beta_{4}$ & -0.8720 & 0.5860 & {[}-2.3496 ; -0.0818{]} \\
$\beta_{5}$ & 2.6536  & 1.7274 & {[}-0.4743 ; 6.4581{]}  \\
$\alpha$    & -0.1207 & 0.0717 & {[}-0.2555 ; -0.0177{]} \\
$\psi$      & 0.6887  & 0.0840 & {[}0.5300 ; 0.8598{]}   \\ \hline
\end{tabular}
\label{tab:inferential_sumamrykgomp}
\end{table}

Table \ref{tab:inferential_sumamrykgomp} shows that the DGGD exhibited a defective form, as evidenced by the negative estimated value of the parameter $\alpha$, with statistical significance shown by the estimated 95\% credibility interval. Based on the regression structure for the cure parameter and the estimated coefficients from the dataset, disregarding the effects of other covariates, the expected probability of cure for patients with uterine cancer, obtained from $\beta_{0}$ applied to the logistic link function, is approximately 12\%. Regarding other covariates, patients aged over 50 years exhibit, on average, a 10\% reduction in the probability of cure, assuming no influence from additional treatments. The positive estimated effect associated with surgery ($\beta_{3} = 2.3608$), along with its corresponding 95\% credibility interval, indicates that surgical intervention has a beneficial and significant impact on increasing the probability of cure. A positive effect was also observed for hormone therapy ($\beta_{5} = 2.6536$), although this result was not statistically significant. In contrast, the occurrence of metastasis, indicated by the presence of distant recurrence, as well as chemotherapy, were identified as risk factors that reduce the probability of cure, given the negative estimated effects of these binary variables and their respective 95\% credibility intervals.

 Table \ref{tab:DIC} presents the values of the model comparison metrics DIC (deviance information criterion), PSIS-LOO (Pareto smoothed importance sampling for Leave-One-Out Cross validation) and -2*LPML (logarithm of pseudomarginal likelihood) for the regression cure models based on the Gompertz and generalized Gompertz distributions. All metrics choose the more appropriate model with the lowest value and they are better described in Appendix \ref{appenxc}. The results indicate that the values are substantially lower for the flexible defective version compared to the traditional form. Therefore, for the uterine cancer dataset, the most appropriate cure model is the one based on the generalized Gompertz distribution. 

\begin{table}[]
\centering
\caption{Values of DIC, PSIS-LOO, -2*LPML for generalized Gompertz and Gompertz defective regression models for uterine cancer dataset.}
\begin{tabular}{lccc} \hline
Model                & DIC     & PSIS-LOO & -2*LPLM  \\ \hline
Generalized Gompertz & 1298.52 & 1327.09   & 1327.15 \\
Gompertz             & 1328.69 & 1331.40   & 1331.59  \\ \hline
\end{tabular}
\label{tab:DIC}
\end{table}

Figure \ref{fig:goodness_of_fit_kreg} presents the PSIS-LOO individual values and the natural logarithm of the CPO values for identifying influential points in the estimation of the generalized Gompertz model. The PSIS-LOO diagnostic indicated that most observations had reliable approximations, with Pareto k values below 0.7. However, one influential observation (212 data point) exceeded the threshold (k > 0.7), suggesting that the leave-one-out predictive distribution for this point is not well-approximated by importance sampling. This indicates this observation is an potential outlier with high leverage. On the other hand, the logarithmic values of the Conditional Predictive Ordinates (CPO) for the generalized Gompertz model were relatively high, indicating higher CPO values and, consequently, good predictive performance. This suggests that the model provides an adequate fit in terms of prediction.

\begin{figure}[htbp]
\centering
\begin{subfigure}[b]{0.35\textwidth}
  \centering
  \includegraphics[width=\linewidth]{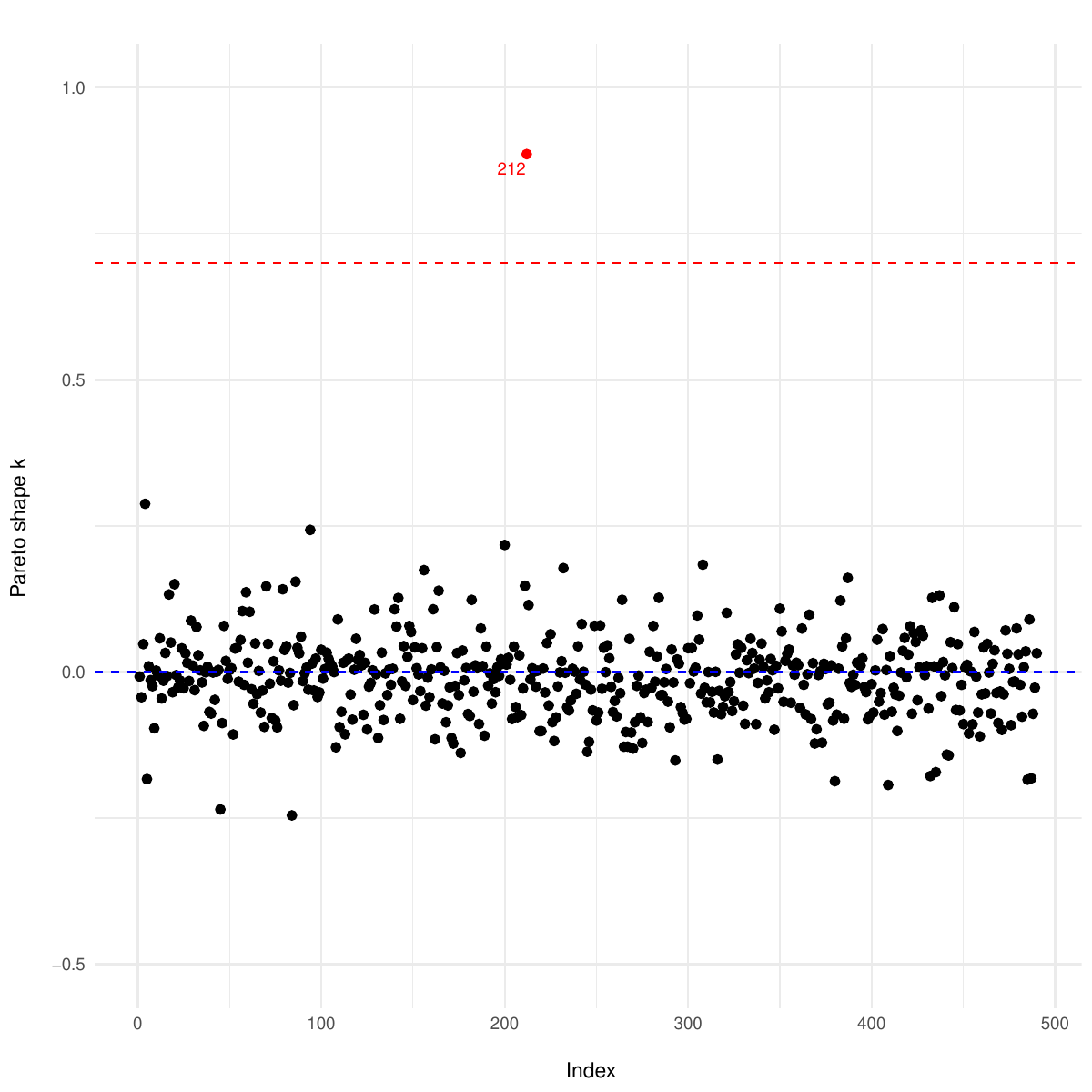}
  \caption{}
  \label{fig:sub1}
\end{subfigure}
\hspace{0.8 cm}
\begin{subfigure}[b]{0.35\textwidth}
  \centering
  \includegraphics[width=\linewidth]{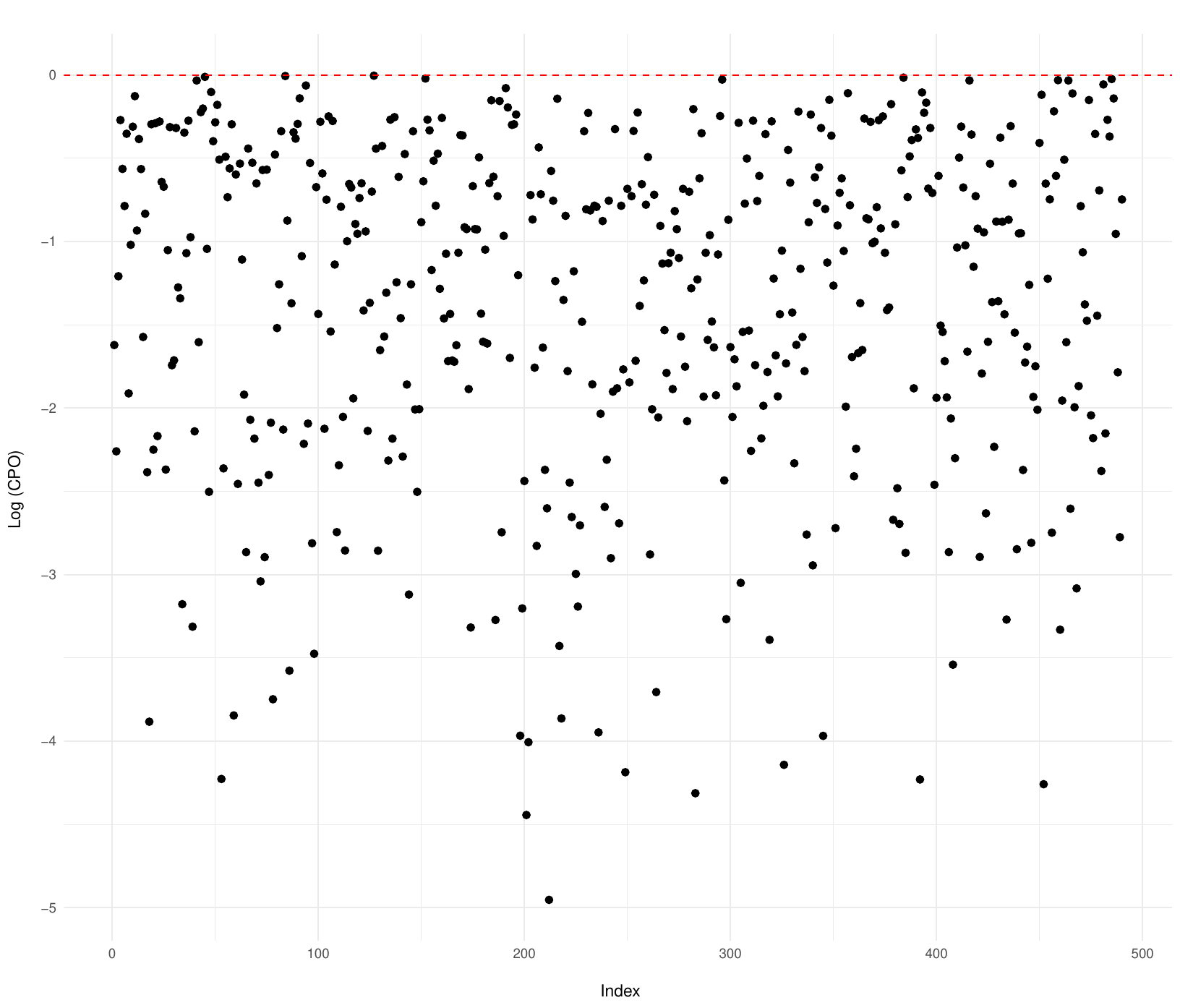}
  \caption{}
  \label{fig:sub2}
\end{subfigure}
\caption{Diagnostic plot of the Pareto shape parameters $k$ obtained from the PSIS-LOO approximation (a), and Conditional Predictive Ordinate (CPO) values (b), based on the DGGD fitted to the uterine cancer dataset.}
\label{fig:goodness_of_fit_kreg}
\end{figure}

Figure \ref{fig:goodness_of_fit_kreg} presents the values of martigales residuals and deviance residuals, which are classical tools to identify outliers. The plots show that the residuals are randomly scattered around zero, indicating that the regression model provides a reasonably good fit to the data. Only one outlier (284 data point) falls outside the range of (–3, 3) in the deviance residuals, which supports the adequacy of the model. Overall, the results provide evidence that the regression model appropriately captures the relationship underlying the cure fraction.

Appendix \ref{appendixA} presents the inferential summaries excluding observations 212 and 284, which were identified as influential and outlier points, respectively. Although the marginal standard deviations of some parameter estimates increased considerably, the statistical significance of most regression coefficients was preserved. When observation 212 was removed, the regression coefficient associated with hormonal treatment for uterine cancer remained statistically significant, while the presence of metastasis did not show a significant effect. Overall, the inferential summary without observation 212 was very similar to that obtained using the complete dataset.

\begin{figure}[htbp]
\centering
\begin{subfigure}[b]{0.35\textwidth}
  \centering
  \includegraphics[width=\linewidth]{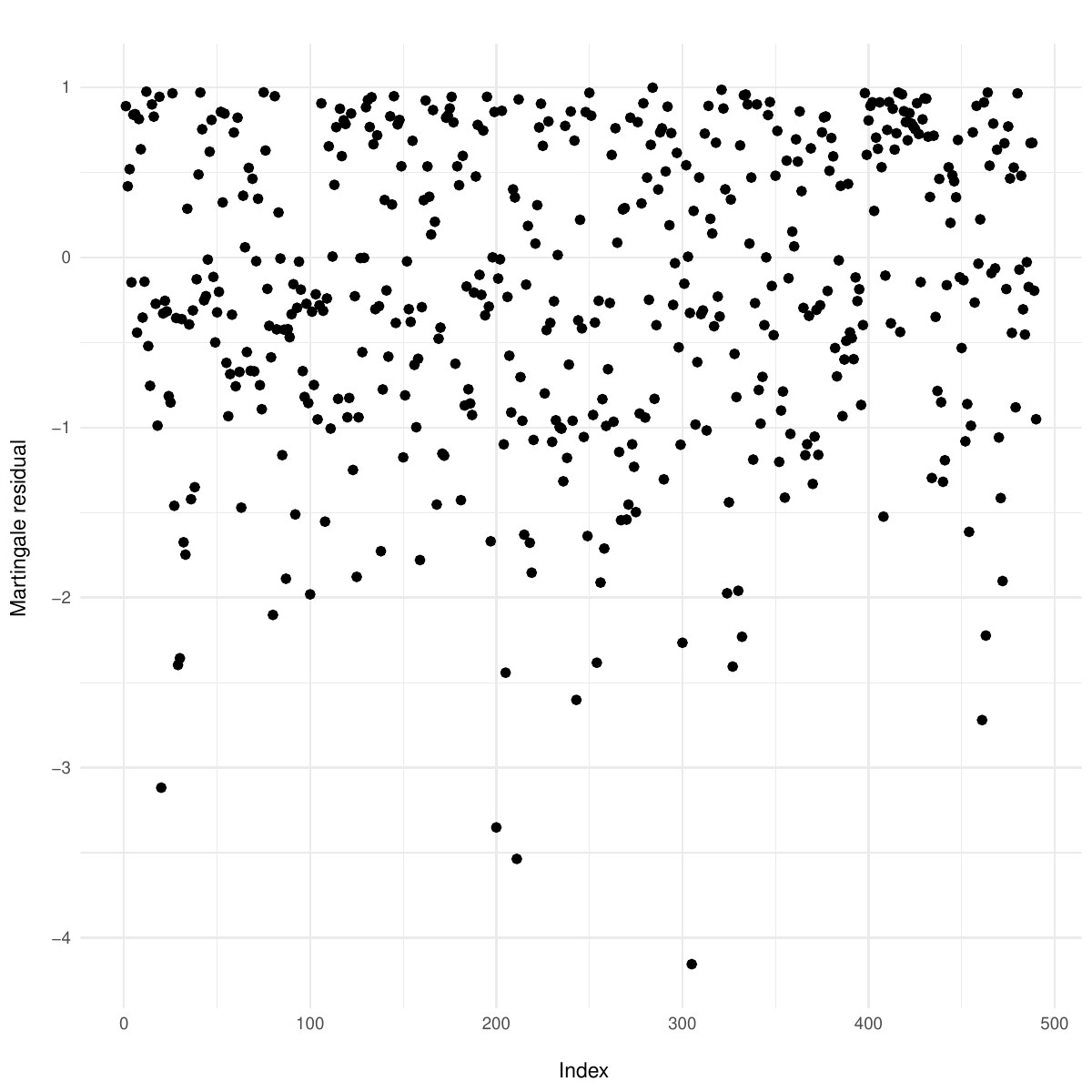}
  \caption{}
  \label{fig:sub1}
\end{subfigure}
\hspace{0.8 cm}
\begin{subfigure}[b]{0.35\textwidth}
  \centering
  \includegraphics[width=\linewidth]{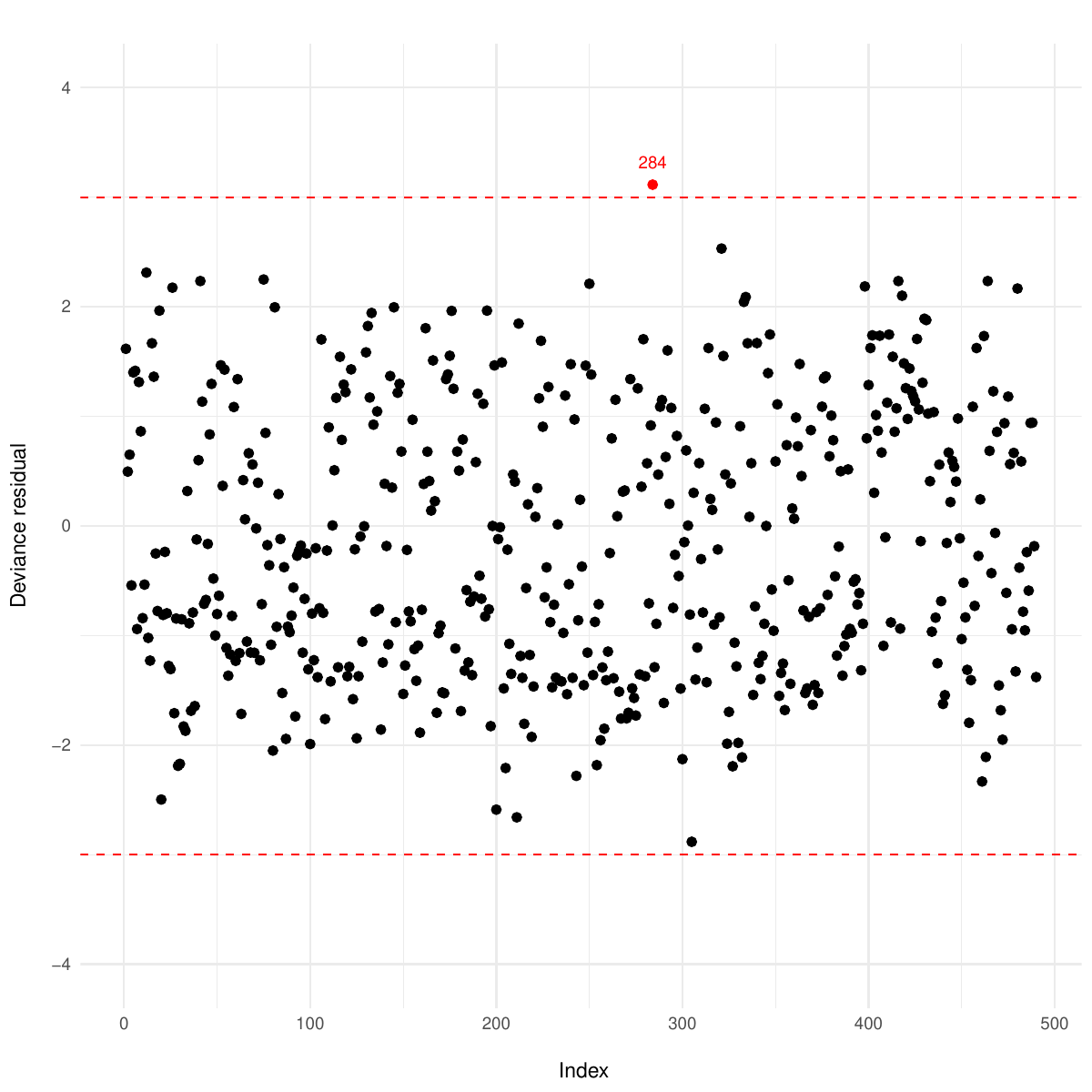}
  \caption{}
  \label{fig:sub2}
\end{subfigure}
\caption{Martingale residuals (a) and deviance residuals (b) based on the defective generalized Gompertz model fitted to the uterine cancer dataset.}
\label{fig:goodness_of_fit_kreg}
\end{figure}

Defective models arise as alternatives to standard mixture models from a parametric perspective. A common approach in the literature is to specify the Weibull distribution for the non-cured individuals in mixed populations. In Appendix \ref{appendixB}, we present the inferential results and diagnostic plots for the standard mixture model assuming a Weibull distribution. The model was constructed similarly to the one developed in this project, where treatments, age, and metastasis served as binary covariates to interpret the patient’s probability of cure, with regression coefficients assigned with vague priors. Upon analysis, the inferential summary revealed high standard deviations for the parameter distributions, and no statistically significant effects were observed for the factors commonly identified in the medical literature as relevant for cancer cure \citep{boggess2020uterine, green2001survival, lee2007prognostic}. Additionally, the PSIS-LOO influential point diagnostics indicated high Pareto $k$ values, with several exceeding the high (0.7) and critical (1.0) thresholds, suggesting the presence of many influential observations. Moreover, many data poitns showed values closed to zero in terms of the  logarithmic values of the CPO, indicating a loss of predictive accuracy in the standard mixture model. Finally, the deviance residuals showed that several data points had values below –3, suggesting the presence of multiple outliers in the regression model.

\section{Conclusion}
\label{sec:conclusion}

In this study, we aimed to demonstrate how hospital-based intervention factors can contribute valuable information regarding the probability of cure for patients with uterine cancer in Brazil. Our Bayesian defective regression model has shown important interpretative results, pointing out the age, presence of metastasis and even the chemotherapy treatment as risk factors to the probability of cure. The effect and significant surgery intervention is highly recommended to increase the probability of immunity over the years to women previously diagnosed with uterine cancer.

Our statistical model for long-term survival data is based on a reparametrization of the defective generalized Gompertz distribution, using its information of cure in terms of the survival function to incorporate relevant clinical information within a regression framework. This approach has not yet been explored in clinical research and can be an interesting alternative to regression model based on the mixture cure formulation. We believe that our Bayesian offers substantial interpretability for the medical field, particularly in the context of diseases with high survival rates driven by advances in medical treatments that gradually enhance patient immunity over time.

We have also demonstrated that our estimation procedure, based on HMC sampling, accurately recovers the true parameter values in simulation studies, while exhibiting key advantages in real-world applications by identifying statistically significant cure-related factors consistent with the medical literature. Furthermore, we discussed how reparametrization in terms of the cure fraction provides meaningful insights into the influence of cure factors in uterine cancer.

Our model achieved a good fit to the real-world dataset and offered a more interpretable alternative compared to traditional mixture cure models commonly based on the Weibull distribution. Additionally, the available dataset includes geographic information on patient location, which could be of particular interest for assessing whether spatial distribution influences the probability of cure. Incorporating spatial structures into the regression framework represents a promising avenue for future research.

\section{Acknowledgments}

This research was supported by FAPESP (\textit{Fundação de Amparo à Pesquisa do Estado de São Paulo}) under grant number 24/07832-6. This reseach was also carried out using the computational resources of the Center for Mathematical Sciences Applied to
Industry (CeMEAI) funded by FAPESP (grant 2013/07375-0).






\newpage

\appendix
\section{Model Estimation of the Defective Generalized Gompertz Distribution After Removing Influential and Outlier Observations}
\label{appendixA}

\begin{table}[H]
\centering
\caption{Estimates values of mean, standard deviation (SD) and 95\% credibility interval for uterine cancer data for the defective generalized Gompertz regression model without influential and outliers observations.}
\begin{tabular}{ccccc} \hline
Removed   data                & Parameter   & Mean    & SD     & C.I (95\%)              \\ \hline
\multirow{8}{*}{\{212\}}      & $\beta_{0}$ & -1.6091 & 1.2607  & {[}-4.5002 ; -0.2314{]} \\
                              & $\beta_{1}$ & -1.5288 & 0.5653  & {[}-2.9047 ; -0.7583{]} \\
                              & $\beta_{2}$ & -1.4016 & 1.2243  & {[}-5.0815 ; 0.4363{]}  \\
                              & $\beta_{3}$ & 2.1081  & 0.9748  & {[}1.0351 ; 4.4031{]}   \\
                              & $\beta_{4}$ & -0.9703 & 0.7653  & {[}-3.0372 ; -0.1130{]} \\
                              & $\beta_{5}$ & 10.0157 & 5.0179  & {[}2.3944 ; 21.5160{]}  \\
                              & $\alpha$    & -0.1281 & 0.0703  & {[}-0.2588 ; -0.0238{]} \\
                              & $\psi$      & 0.6960  & 0.0822  & {[}0.5558 ; 0.8646{]}   \\ \hline
\multirow{8}{*}{\{284\}}      & $\beta_{0}$ & -1.5400 & 1.1759  & {[}-4.6326 ; -0.2403{]} \\
                              & $\beta_{1}$ & -1.5118 & 0.5969  & {[}-3.0494 ; -0.7228{]} \\
                              & $\beta_{2}$ & -1.8244 & 1.4747  & {[}-5.5468 ; -0.0384{]} \\
                              & $\beta_{3}$ & 2.0526  & 0.8777  & {[}1.0695 ; 4.1899{]}   \\
                              & $\beta_{4}$ & -0.8103 & 0.5315  & {[}-1.9640 ; -0.0854{]} \\
                              & $\beta_{5}$ & 2.4541  & 1.7977  & {[}-1.0174 ; 6.1314{]}  \\
                              & $\alpha$    & -0.1359 & 0.0730  & {[}-0.2706 ; -0.0258{]} \\
                              & $\psi$      & 0.7171  & 0.0868  & {[}0.5629 ; 0.8944{]}   \\ \hline
\multirow{8}{*}{\{212, 284\}} & $\beta_{0}$ & -1.0921 & 0.6910  & {[}-2.9899 ; -0.1768{]} \\
                              & $\beta_{1}$ & -1.3656 & 0.3739  & {[}-2.2627 ; -0.7602{]} \\
                              & $\beta_{2}$ & -1.2823 & 0.7954  & {[}-3.2151 ; 0.0033{]}  \\
                              & $\beta_{3}$ & 1.7447  & 0.5108  & {[}1.0293 ; 3.0791{]}   \\
                              & $\beta_{4}$ & -0.6979 & 0.3847  & {[}-1.6592 ; -0.0896{]} \\
                              & $\beta_{5}$ & 9.7128  & 5.1857  & {[}2.2278 ; 21.5396{]}  \\
                              & $\alpha$    & -0.1533 & 0.0608  & {[}-0.2700 ; -0.0431{]} \\
                              & $\psi$      & 0.7342  & 0.0780  & {[}0.5959 ; 0.8965{]}   \\ \hline
\end{tabular}
\end{table}

\newpage
\section{Model Estimation and Diagnostic Assessment for the Weibull Cure Rate Model}
\label{appendixB}

\begin{table}[H]
\centering
\caption{Estimates values of mean, standard deviation (SD) and 95\% credibility interval for uterine cancer data of the Weibull mixture cure model, with shape $\lambda$ and rate $\gamma$ parameters.}
\begin{tabular}{cccc} \hline
Parameter   & Mean    & SD     & C.I (95\%)             \\ \hline
$\beta_{0}$ & 14.4322 & 6.2324 & {[}4.6332 ; 28.1547{]}   \\
$\beta_{1}$ & 5.1562  & 7.5016 & {[}-8.5172 ; 20.8941{]}  \\
$\beta_{2}$ & 2.2739  & 8.6438 & {[}-13.6819 ; 20.2116{]} \\
$\beta_{3}$ & 4.9779  & 7.6339 & {[}-8.7768 ; 21.1380{]}  \\
$\beta_{4}$ & 4.8229  & 7.8081 & {[}-9.4174 ; 21.3307{]}  \\
$\beta_{5}$ & 2.6863  & 8.3688 & {[}-12.1063 ; 20.6566{]} \\
$\lambda$   & 1.5842  & 0.1234 & {[}1.3549 ; 1.8369{]}     \\
$\gamma$    & 0.8337  & 0.0400 & {[}0.7568 ; 0.9130{]}     \\ \hline
\end{tabular}
\end{table}

\begin{figure}[htbp]
\centering
\begin{subfigure}[b]{0.35\textwidth}
  \centering
  \includegraphics[width=\linewidth]{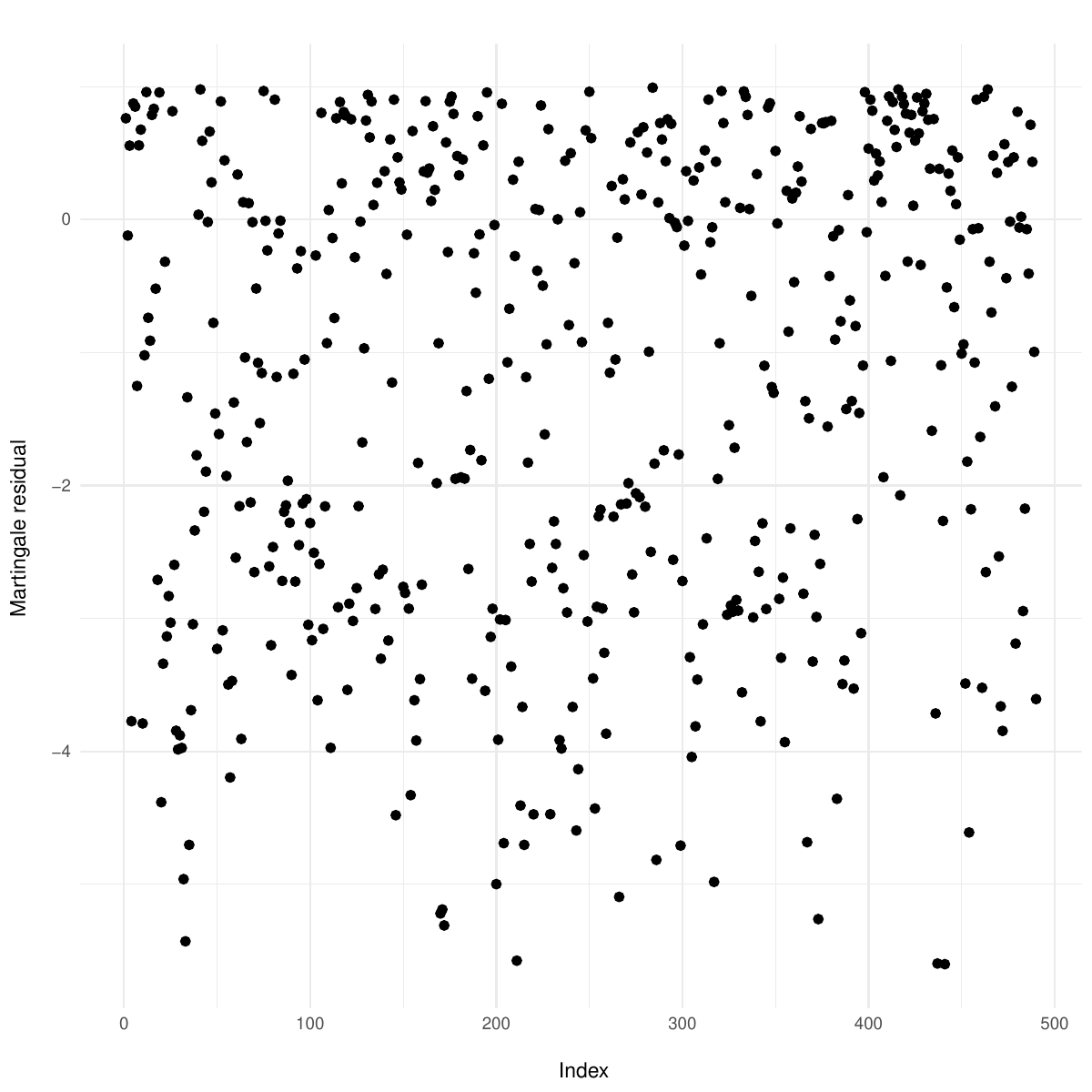}
  \caption{}
  \label{fig:sub11}
\end{subfigure}
\hspace{0.8 cm}
\begin{subfigure}[b]{0.35\textwidth}
  \centering
  \includegraphics[width=\linewidth]{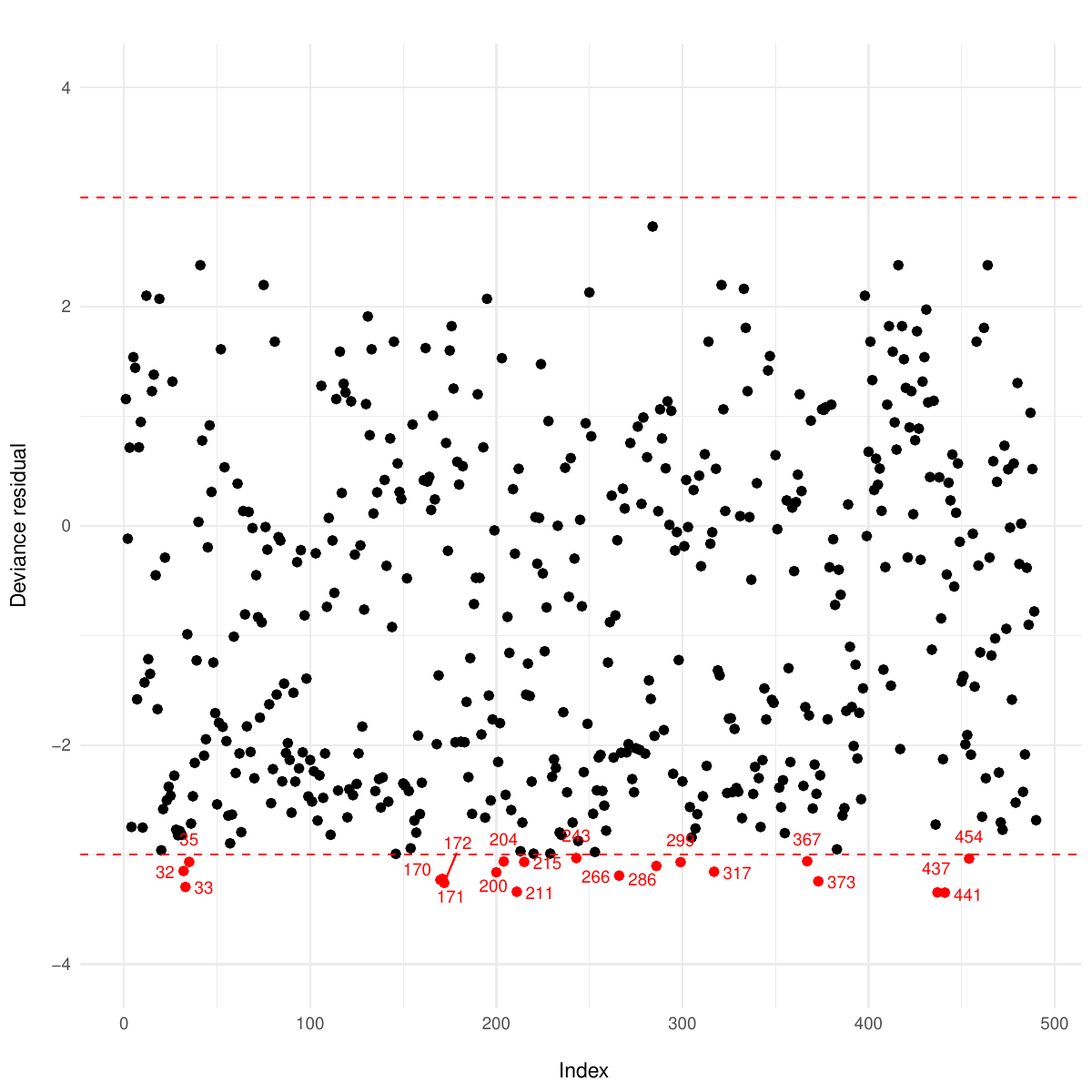}
  \caption{}
  \label{fig:sub22}
\end{subfigure}
\caption{Martingale residuals (a) and deviance residuals (b) based on the Weibull mixture cure model fitted to the uterine cancer dataset.}
\label{fig:goodness_of_fit_kreg}
\end{figure}

\begin{figure}[htbp]
\centering
\begin{subfigure}[b]{0.35\textwidth}
  \centering
  \includegraphics[width=\linewidth]{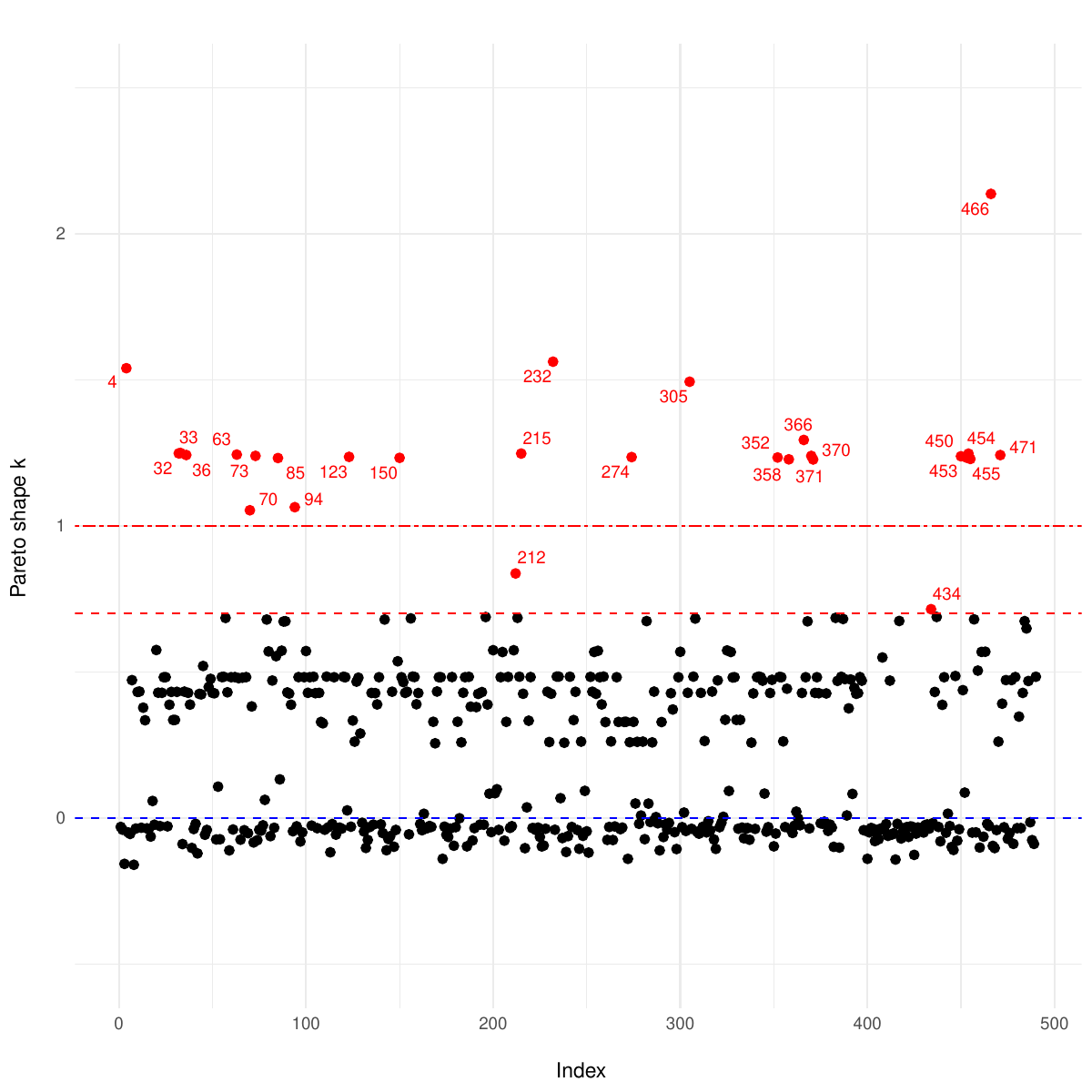}
  \caption{}
  \label{fig:sub1}
\end{subfigure}
\hspace{0.8 cm}
\begin{subfigure}[b]{0.35\textwidth}
  \centering
  \includegraphics[width=\linewidth]{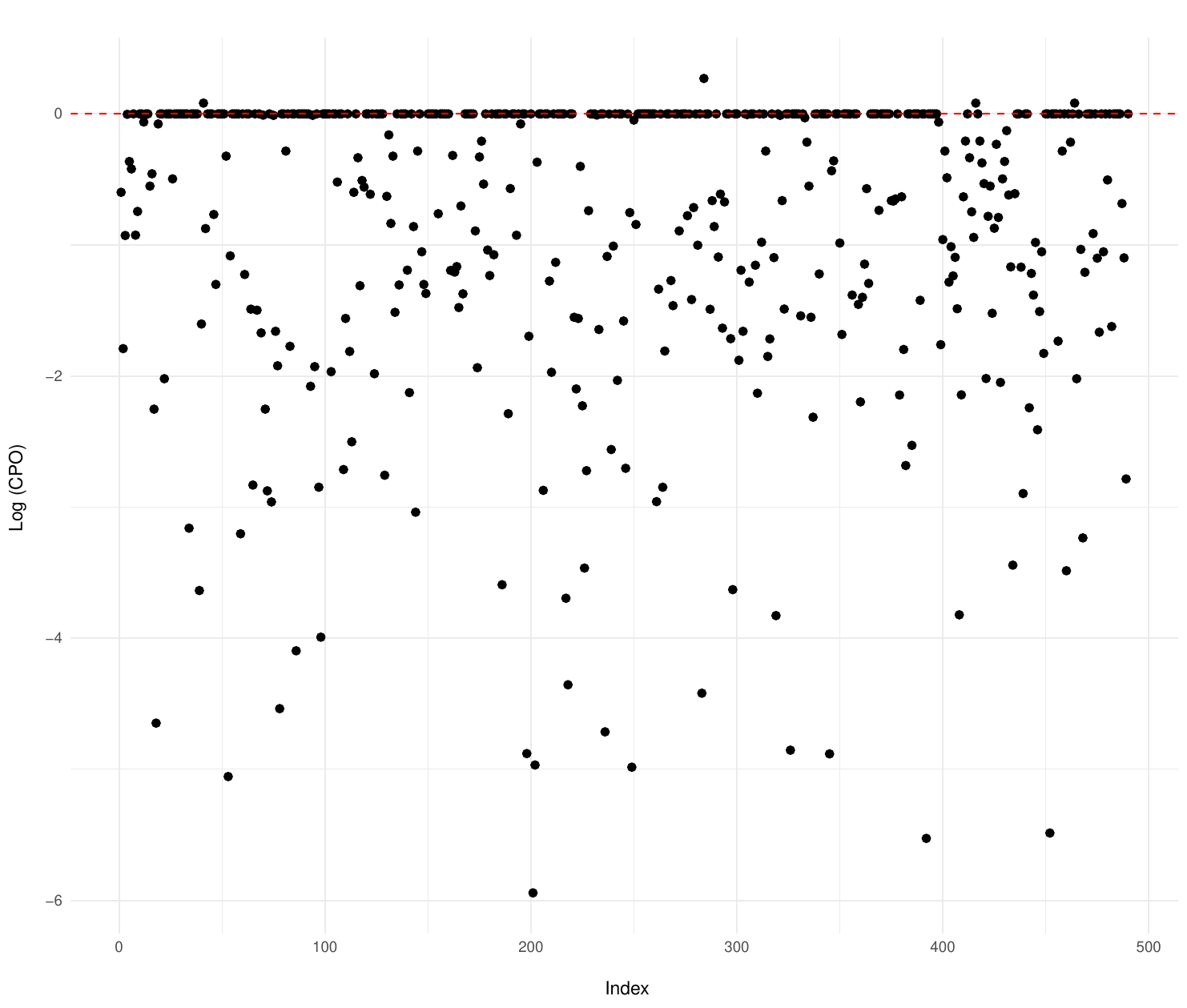}
  \caption{}
  \label{fig:sub2}
\end{subfigure}
\caption{Diagnostic plot of the Pareto shape parameters $k$ obtained from the PSIS-LOO approximation (a), and logarithm of Conditional Predictive Ordinates (CPO) values (b), based on the Weibull mixture cure model fitted to the uterine cancer dataset.}
\label{fig:goodness_of_fit_kreg}
\end{figure}

\section{Model Selection Criteria in Bayesian Models}
\label{appenxc}

Let $y_{-i}$ denote the complete dataset except for the $i-$observation ($i=1,\dots,n$), the CPO for the $i-$th observation ($\mathrm{CPO}_i$) emerges as the posterior harmonic mean of the $i-$th observed likelihood:

\begin{align}
\mathrm{CPO}_i &= p(y_i \mid y_{(-i)}) = \int p(y_i \mid \boldsymbol{\vartheta}) \, p(\boldsymbol{\vartheta} \mid y_{(-i)}) \, d\boldsymbol{\vartheta} \nonumber \\
&= \int p(y_i \mid \boldsymbol{\vartheta}) \left[ \frac{ \prod_{j \neq i} p(y_j \mid \boldsymbol{\vartheta}) \, p(\boldsymbol{\vartheta}) }{ \int \prod_{j \neq i} p(y_j \mid \boldsymbol{\vartheta}) \, p(\boldsymbol{\vartheta}) \, d\boldsymbol{\vartheta} } \right] d\boldsymbol{\vartheta} \nonumber \\
&= \frac{ p(y) }{ \int \prod_{j \neq i} p(y_j \mid \boldsymbol{\vartheta}) \, p(\boldsymbol{\vartheta}) \, d\boldsymbol{\vartheta} } 
\int p(y \mid \boldsymbol{\vartheta}) \, p(\boldsymbol{\vartheta}) \, d\boldsymbol{\vartheta} \nonumber \\
&= \left( \int \frac{ \prod_{j \neq i} p(y_j \mid \boldsymbol{\vartheta}) \, p(\boldsymbol{\vartheta}) }{ p(y) } \, d\boldsymbol{\vartheta} \right)^{-1}\nonumber \\
&= \left( \int \frac{ p(y_i \mid \boldsymbol{\vartheta}) \prod_{j \neq i} p(y_j \mid \boldsymbol{\vartheta}) \, p(\boldsymbol{\vartheta}) }{ p(y_i \mid \boldsymbol{\vartheta}) \, p(y_{(-i)}) } \, d\boldsymbol{\vartheta} \right)^{-1} \nonumber \\
&= \left( \int \frac{ 1 }{ p(y_i \mid \boldsymbol{\vartheta}) } \, p(\boldsymbol{\vartheta} \mid y) \, d\boldsymbol{\vartheta} \right)^{-1}.
\label{eq:cpo}
\end{align}

The Equation \ref{eq:cpo} shows that it is possible to perform cross validation without a separate analyses. Using MCMC samples, we can estimate the individual CPO as

\begin{equation}
    \widehat{\mathrm{CPO}_i} = \left(\frac{1}{n}\sum_{k=1}^{n} \dfrac{1}{p(y_{i} \mid \boldsymbol{\vartheta}_{k})}\right)^{-1},
\end{equation}

where $\boldsymbol{\vartheta}_{k}$ is the $k-$th sampled parameter vector from an MCMC analysis from the sampled posterior distribution ($k=1, \dots, K$) [citar Chen 2000]. The quantity $p(y_{i} \mid \boldsymbol{\vartheta}_{k})$ is generally only available on the log scale, therefore, we have the convenient formula

\begin{align}
    \log \left(\widehat{\mathrm{CPO}_i}\right) = \log(n) + l_{i, min} - \log \left(\sum_{k=1}^{n} \exp\{l_{i, min} - l_{i,k}\}\right),
\end{align}

where $l_{i,k} = \log(p(y_{i} \mid \boldsymbol{\vartheta}_{k}))$ is the observed likelihood for the $i-$th observation using the $k-$th sampled parameter vector, and $l_{i,min} = min\{l_{i,k} : k= 1, \dots, n\}$.

The logarithm of pseudomarginal likelihood (LPML) is estimated by sum of the individual CPO's values:

\begin{equation}
    \mathrm{LPML} = \sum_{i=1}^{n} \log \left(\widehat{\mathrm{CPO}_i}\right).
\end{equation}

The Deviance Information Criterion (DIC) is another measure to compare Bayesian models, which penalizes the model fitting based on complexity \citep{spiegelhalter2002bayesian}. The DIC can be expressed as

\begin{equation}
    D(\boldsymbol{\vartheta}) = -2 \, \sum_{i=1}^{n} \log \left(f(y_{i} \mid \boldsymbol{\vartheta} )\right),
\end{equation}

\noindent where $f(.)$ is the probability density function of the defective model under evaluation. The posterior mean of $D(\boldsymbol{\vartheta})$ can be estimated from samples of posterior distribution. Indeed, $\bar{D} = \sum_{k=1}^{K} D(\boldsymbol{\vartheta}_{k})$, where the $\boldsymbol{\vartheta}_{k}$ is the $k-$th posterior sample. According to \cite{spiegelhalter2002bayesian}, the DIC is estimated as

\begin{equation}
    \widehat{DIC} = \bar{D}  + \frac{1}{2} p_{D},
\end{equation}

where $p_{D} = \bar{D} - D(\boldsymbol{\bar{\vartheta}})$ is the effective parameters numbers, and $D(\bar{\boldsymbol{\vartheta}}) = -2 \log(p(y \mid \bar{\boldsymbol{\vartheta}}))$ is the deviance evaluated at the posterior mean of the parameters, where 
$\bar{\boldsymbol{\vartheta}}$ is typically the posterior mean vector of parameters.

To improve the accuracy and stability of leave-one-out (LOO) cross-validation estimates, we employ Pareto Smoothed Importance Sampling (PSIS) as proposed by \citep{vehtari2015efficient}. This method regularizes the distribution of importance sampling weights by smoothing extreme values that often arise in Bayesian models, particularly when importance ratios exhibit heavy-tailed behavior.

Importance sampling is known to be sensitive to extreme weights, which can result in unstable and high-variance estimates. PSIS mitigates this issue by fitting a generalized Pareto distribution (GPD) to the upper tail of the distribution, specifically, the top 20\% of the importance ratios. The fitted shape parameter $\hat{k}$ serves as a diagnostic measure for assessing the reliability of the estimates. This approach builds on and extends earlier diagnostic tools developed by \cite{peruggia1997variability} and \cite{epifani2008case}.

To obtain the PSIS values for each individual, we follows the steps:

\begin{itemize}
    \item [($i$)] For each data point $i$, $i =1 , \dots, n$, a GPD is fitted to the largest 20\% of the importance weights 
    $$r_{i}^{s} = \dfrac{1}{p(y_{i} \mid \boldsymbol{\vartheta}^{s})} \varpropto \dfrac{p(\boldsymbol{\vartheta}^{s} \mid y_{-i})}{p(\boldsymbol{\vartheta}^{s} \mid y)}.$$
    This is done independently for each observation using empirical Bayes estimation; \\
    \item [($ii$)] The $M$ largest weights are replaced by their expected order statistics from the fitted GPD, using the inverse cumulative distribution function
  $$
  F^{-1}\left( \frac{z - 1/2}{M} \right), \quad z = 1, \ldots, M.
  $$

  where $M$ is the number of simulation draws used to fit the Pareto (in this case, $M = 0.2 S$). This results in a smoothed set of weights $\tilde{w}_s^i$, indexed by simulation draws $s$ and data points $i$; \\
   \item [($iii$)] To guarantee finite variance of the estimate, the stabilized weights are truncated at $S^{3/4} \bar{w}_i$, where $\bar{w}_i$ is the average of the smoothed weights for observation $i$. The final truncated weights are denoted by $w_s^i$.
\end{itemize}

The above steps are repeated for each observation ($i = 1, \dots, n$), to get the vector of weights $w_{i}^{s}, s = 1, \dots, S$, for each $i = 1, \dots, n$. The weights should present a better behavior than the raw importance rations $r_{i}^{s}$.

The results can then be combined to compute the desired LOO estimates. The PSIS is based on the -2 times the value of the expected log pointwise predictive density given by 

\begin{equation}
\widehat{\text{elpd}}_{\text{psis-loo}} = \sum_{i=1}^{n} \log \left( 
\frac{ \sum_{s=1}^{S} w_i^s p(y_i \mid \theta^s) }{ \sum_{s=1}^{S} w_i^s }
\right).
\end{equation}

For the individual values of $\hat{k}$ of the generalized Pareto distribution, \cite{vehtari2015efficient} guarantee reliable estimates under the central limit theorem for $\hat{k}$. If the estimated tail shape parameter $\hat{k}$ exceeds 0.5 but is less than 0.7, the estimation the generalized central limit theorem for stable distributions holds. If $\hat{k} > 0.7$, the approximation is not reliable and it is an evidence of influential data point in the Bayesian model. If $\hat{k} > 1.0$, the variance and the mean of the raw ratios distribution do not exist and it affects directly the estimation of PSIS-LOO, indicating poor predictive performance \citep{vehtari2015efficient}.

\end{document}